\documentclass[twocolumn,pre,amsmath,amssymb,aps,nofootinbib,superscriptaddress,showpacs,floatfix]{revtex4-2}
\usepackage{graphicx}
\usepackage{bm}
\usepackage{physics}
\usepackage{upgreek}
\usepackage{xcolor}
\usepackage{tabularx}
\usepackage{booktabs}

\let\vec\mathbf

\newcommand\me{\mathrm{e}}

\renewcommand*\grad[1]{\nabla{#1}}
\renewcommand*\div[1]{\nabla\cdot\vec{#1}}
\renewcommand*\curl[1]{\nabla\times\vec{#1}}

\newcommand*\laps[1]{\Delta^*{#1}}
\newcommand*\mdv[2]{\left(\vec{#1}\cdot\nabla\right)\vec{#2}}

\newcommand{\grf}{\grad{\phi}}

\newcommand{\grp}{\grad{\psi}}
\newcommand{\tb}[1]{_\text{#1}}

\newcommand{\ie}{\emph{i.e., }}

\newcommand{\reff}[1]{(\ref{#1})}
\newcommand{\eref}[1]{Eq.\reff{#1}}

\newcolumntype{Y}{>{\centering\arraybackslash}X}

\begin{document}

\title{Influence of rotation on axisymmetric plasma equilibria: double-null DTT scenario}

\author{Matteo Del Prete}
\affiliation{ Physics Department, ``Sapienza'' University of Rome,\\
              P.le Aldo Moro 5, 00185 Roma, Italy}

\author{Giovanni Montani}
\affiliation{ ENEA, Fusion and Nuclear Safety Department, C. R. Frascati,\\
              Via E. Fermi 45, 00044 Frascati (Roma), Italy}
\affiliation{ Physics Department, ``Sapienza'' University of Rome,\\
              P.le Aldo Moro 5, 00185 Roma, Italy}

%\date{\today}

\begin{abstract}
%We solve the Grad--Shafranov equation (GSE) for the equilibrium of a rotating axisymmetric plasma, providing two different analytic expressions for the homogeneous solution. They are respectively suitable when the separatrix shape is described in terms of a few parameters (e.g., the ratio between minor and major radius, the plasma elongation and triangularity), and when the plasma boundary is assigned as an analytic or numerical curve. We use both solutions to describe the double-null plasma scenario at $I\tb{P}=5\,$MA predicted for the upcoming DTT experiment, focusing our attention on how the plasma rotation influences the main features of the equilibrium.
%We study the dependence of some relevant tokamak equilibrium quantities on the toroidal plasma rotation. The Grad--Shafranov equation generalised to the rotating case is analytically solved employing two different representations for the homogenous solution. An expression in terms of polynomials is suitable when the separatrix shape is described by a few geometrical parameters, while an expression in terms of Bessel functions is suitable when the full plasma boundary curve is assigned. We implement both solutions in the realistic double-null plasma scenario predicted for the DTT experiment currently under construction, highlighting the effects of the introduction of a rotation profile on the equilibrium properties. In general cases, In the specific case of the double-null scenario predicted for the DTT experiment, we prove how a variety of rotation profiles are compatible with the machine parameters and 
We study the dependence of some relevant tokamak equilibrium quantities on the toroidal plasma rotation. The Grad--Shafranov equation generalised to the rotating case is analytically solved employing two different representations for the homogenous solution. Using an expression in terms of polynomials, we describe the separatrix shape by a few geometrical parameters, reproducing different plasma scenarios such as double-null and inverse triangularity. In this setting, the introduction of toroidal rotation corresponds to variations on relevant plasma quantities, most notably an enhancement of the poloidal beta. Using a more general expression in terms of Bessel functions, we reconstruct the full plasma boundary of the double-null configuration proposed for the upcoming DTT experiment, demonstrating how said configuration is compatible with different values of the plasma velocity. 
\end{abstract}

\maketitle

%%%%%%%%%%%%%%%%%%%%%%%%%%%%%%%%%%
%%	introduction		%%%%%%%%%%%%%%%%%%%%%%
%%%%%%%%%%%%%%%%%%%%%%%%%%%%%%%%%%
\section{introduction}
The basic concept at the ground of any operational regime of a tokamak device \cite{wesson} is the theoretical existence of an axisymmetric plasma equilibrium \cite{biskamp}. In a real machine, this equilibrium can exist for a time which is inherently limited by the duration of the discharge. The duration is usually much longer than the characteristic timescale on which magnetohydrodynamic instabilities develop, leading to abrupt losses of confinement, and much shorter than dissipation timescales due to resistivity or other non ideal effects, leading to slow losses of confinement.

The description of a tokamak equilibrium is based on the balance of the ideal MHD forces, \ie pressure gradients versus magnetic pressure and tension, resulting in the well-known Grad-Shafranov equation (GSE) \cite{shafranov}, in which the presence of steady matter flux is neglected. This assumption can be motivated by the conditions of operation of tokamak machines, which discharge is, in general, associated to a flux-free quasi-ideal plasma. Nonetheless, the emergence of a \emph{spontaneous rotation} in Tokamak devices has been observed since the early nineties \cite{hassam93,duval07}, both in the toroidal and poloidal directions. Many proposals have been argued in order explain this phenomenon, which can be interpreted as a result of a self--organization of the plasma in the transition from turbulent to laminar flow. Indeed, the transition between turbulent and laminar regimes is an interchange phenomenon, due to the unavoidable linear and nonlinear instability of the rotating plasma. %Hence, predator--prey paradigms can be formulated naturally \cite{...}.

Another important operation condition of a tokamak leading to important rotation profiles is the heating of the plasma via hot neutral beam injection: the beam injected in the tangential direction, tranferring angular momentum into the plasma, can trigger rotation flows inside the configuration \cite{karpushov17}.

According to these considerations, the inclusion of rotation in the computation of a tokamak equilibrium is a relevant topic that may require increasing attention in the years to come. The theory of rotating tokamak equilibria has been developed by many authors (e.g. see \cite{mp1980,hameiri83} and citing articles), with most studies mainly relying on the introduction of a Bernoulli--like function (as in traditional fluid dynamics) in order to generalize the GSE to a plasma with flow, while keeping its mathematical form mostly intact \cite{ogilvie97}.

Here, we investigate the case of a tokamak equilibrium in the presence of a toroidal velocity field and we address its description through the introduction of a generalized pressure function, as in \cite{mp1980}. We first construct simple semi--analytical solutions of the obtained equation, generalizing the well-known Solov'ev scenario \cite{solo68}. Then, we implement our model to study how the double--null configuration at $5\,$MA of the DTT Italian proposal \cite{dttgreen} is modified by the presence of toroidal rotation. We are able to characterize the dependence of some basic plasma quantities on the toroidal velocity, such as the poloidal beta $\beta\tb{pol}$, the plasma current $I\tb{p}$, the profile of the safety factor $q$, the position of the magnetic axis and the morphology of the separatrix with respect to the isobar surface at zero pressure, taken as the plasma boundary. 

Clearly, the introduction of a toroidal rotational field in a Tokamak equilibrium can also be studied via a performing numerical code, see for instance \cite{gbmk2004}. However, our semi--analytical study, based on the construction of a generic solution for a linear GSE, is a powerful tool to establish precise relations among the model parameters. The choice of a linear equilibrium allows to individualize the basic eigenfunctons of the configurational problem and it is justified by the expansion of the unkwnon functions of the magnetic flux function up to the lowest order of approximation.
In this respect, our correlation between the parameter governing the rotation intensity and the $\beta$ value of the plasma must be regarded as a general feature of the considered family of plasma configurations.

The manuscript is structured as follows. In section \ref{sec-eqs}, we recall the fundamental equations from the known literature, we outline the basis for our study introducing the necessary assumptions, and we provide a convenient form for the particular solution of the equilibrium. In section \ref{sec-poly}, we solve the homogeneous problem using a purely polynomial expansion of the magnetic flux function $\psi$. We show how this simple solution is able to represent different plasma scenarios, characterized only by few constraints, and what is the impact of plasma rotation on the equilibrium properties. In section \ref{sec-bessel}, we provide a different, more general solution to the homogeneous problem, which allows for a more precise determination of the plasma separatrix while still mantaining a flexible fitting procedure. We study the specific case of the DTT double-null scenario, illustrating the fitting procedure and the impact of rotation on some relevant equilibrium properties. Concluding remarks follow.

%%%%%%%%%%%%%%%%%%%%%%%%%%%%%%%%%%
%%	equations		%%%%%%%%%%%%%%%%%%%%%%
%%%%%%%%%%%%%%%%%%%%%%%%%%%%%%%%%%
\section{basic equations}\label{sec-eqs}
The equilibrium of magnetically confined plasmas can be described by few basic equations:
\begin{align}
	\rho \mdv{v}{v} = -\grad{P}+\vec{J}\times\vec{B}\,, \label{eq:1}\\
	\div{B} = 0 \,,\\
	\upmu_0\vec{J} = \curl{B} \,,
\end{align}
which express the conservation of the momentum of a charged fluid with density $\rho$, velocity field $\vec{v}$ and pressure $P$ in the presence of self-consistent current density $\vec{J}$ and magnetic field $\vec{B}$. Working in cylindrical coordinates $(R,\phi,Z)$ and assuming axisymmetry, \ie $\partial_\phi f=0$ for any quantity $f$, the magnetic field can be expressed as $\vec{B}=\upmu_0I\grf+\grp\times\grf$, in terms of the two scalar functions $\psi$ and $I$, which are related to the magnetic flux and toroidal magnetic field inside the plasma, respectively. Furthermore in the static case, $\vec{v}=0$, the equilibrium problem reduces to the well-known Grad-Shafranov equation:
\begin{equation}\label{gse}
	\laps{\psi} = -\upmu_0^2 II' -\upmu_0 P' R^2\,,
\end{equation}
where $\laps{}\equiv\partial_R^2-\partial_R/R+\partial_Z^2$, and the prime denotes differentiation of the arbitrary functions $I$ and $P$ with respect to $\psi$, the fundamental degree of freedom of the system. The solutions of this equation for a confined plasma correspond to nested tori of constant $\psi$, and have been extensively studied in the literature [refs]. Earliest analytical studies focus on the Solov'ev scenario, in which the right-hand side of the equation is made independent on $\psi$ by the assumptions:
\begin{align}
	&\upmu_0 P' = S_1\,,\quad \upmu_0^2II' = S_2\,, \\
	&\implies \laps{\psi} = -S_2 -S_1 R^2\,,
\end{align}
with $S_{1,2}=const$. Other choices can be made while still preserving the linearity of the equation, like quadratic source function scenario with $P',II'\sim\psi$, or the dissimilar source function scenario, with $P' \sim const.,\,II'\sim\psi$.\newline

The analysis is more subtle in the case of a plasma configuration rotating in the toroidal direction with velocity $\vec{v} = \omega R^2 \grf$. In this case \eref{gse} can be generalised as
\begin{equation}\label{gserot1}
	\left(\laps{\psi}+\upmu_0^2 II'\right)\grp =
	-\upmu_0 R^2\left(\grad{P}+ \rho R \omega^2 \grad{R}\right)\,,
\end{equation}
where two difficulties arise: the plasma density enters the equilibrium balance, and the pressure is no longer a pure function of $\psi$. However, it is clear from ideal Ohm's law, $\vec{E}+\vec{v}\times\vec{B}=0$, combined with stationary Faraday's law, $\curl{E}=0$, that the rotation frequency $\omega$ is a new surface function, a result also known as corotation theorem \cite{ferraro}. The set of equations must be closed introducing an equation of state for the fluid, with many possible choices \cite{guazz21}; here we consider the ideal gas law $P=\rho k T$, where $k$ is the Boltzmann constant divided by the ion mass and $T$ is the plasma temperature, which can be safely assumed to be a surface function in tokamak equilibrium configurations, due to the high parallel transport in these devices.
In view of these assumptions, \eref{gserot1} can be rewritten introducing an auxiliary function $\theta(\psi)$ as:
\begin{align}
	&\theta(\psi) \equiv kT\log\frac{\rho}{\rho_0}-\frac{\omega^2 R^2}{2} \,, \\
	&\laps{\psi} = -\upmu_0^2 II' - \upmu_0 R^2 \rho
	\left[ \theta' + R^2\omega\omega' +\left(1-\log\frac{\rho}{\rho_0} \right) kT' \right] \,.
\end{align}
This expression is further simplified defining a generalized pressure $P\tb{T}(\psi)=\rho_0kT\exp(\theta/kT)=P \exp(-\omega^2R^2/2kT)$ which is a source function itself, and coincides with the thermodynamic pressure in the $\omega\to0$ limit \cite{mp1980}. Finally we have:
\begin{equation}\label{eq:fullrot}
	\laps{\psi} = -\upmu_0^2 II' - \upmu_0 R^2 
	\left[ P\tb{T}' + P\tb{T} R^2 \left( \frac{\omega^2}{2kT} \right)'\right]
	e^{\frac{\omega^2R^2}{2kT}} \,,
\end{equation}
which gives the equilibrium of a rotating plasma once the arbitrary functions $I$, $P\tb{T}$, $\omega$ and $T$ are assigned.

Before continuing our analysis, let us introduce the following normalizations: defining the plasma major radius as $R_0$ and the toroidal magnetic field at the major radius as $B_0$, we normalize length with $R_0$, magnetic field with $B_0$, magnetic flux with $B_0R_0^2$, pressure with $B_0^2/2\upmu_0$, $I$ with $B_0R_0/\upmu_0$ and current density with $B_0/\upmu_0R_0$. All quantities are to be meant adimensional from now on, unless stated otherwise.

\subsection*{Solov'ev-like configuration}
Similarly to the Solov'ev assumption in the static scenario, we can make the right-hand side of Eq.(\ref{eq:fullrot}) independent on $\psi$ by setting
\begin{equation}
	\frac{P\tb{T}'}{2} = P_1 \,, \quad 
	II' = I_1 \,, \quad
	\frac{\omega^2R_0^2}{2kT}=M^2 \,,
\end{equation}
where $P_1,I_1,M$ are assumed as contants. The latter expresses the ratio of plasma velocity to thermal velocity at the plasma major radius, and it serves as a parameter to introduce rotation in the equilibrium computation.
The resulting expressions for the equilibrium equation and the relevant quantities are:
\begin{align}
	\laps{\psi} = -R^2 P_1 e^{M^2R^2} - I_1 \,, \\
	P\tb{T}(\psi) = 2P_1\psi\,,\quad P(\psi,R)=2P_1\psi e^{M^2R^2} \,, \label{eqpressure} \\
	I(\psi) = \sqrt{2I_1\psi+I_0}\,,\quad \omega(\psi) = \frac{M}{R_0}\sqrt{2kT(\psi)} \,, \label{eqi}
\end{align}
where $I_0$ is an integration constant introduced to take into account the vacuum toroidal magnetic field.
As usual in the theory of linear differential equations, the full solution is given by the sum of a particular solution $\psi\tb{P}$, plus the general homogeneous solution defined by $\laps{\psi\tb{H}}=0$. It is easy to verify by substitution that the former can be written as:
\begin{equation}\label{eqpart}
	\psi\tb{P} = \frac{P_1}{4M^4}\left[1+M^2R^2-e^{M^2R^2} \right] -\frac{I_1}{2}Z^2\,.
\end{equation}
In this form, we naturally recover the static Solov'ev solution $-P_1R^4/8-I_1Z^2/2$ in the $M\to0$ limit.

%\subsection*{Dissimilar functions configuration}
%Analytical solutions can also be found  with different assumptions on the source functions. The dissimilar functions approach consists in setting
%\begin{equation}
%	\frac{P\tb{T}'}{2} = P_1 \,, \quad 
%	II' = I_1^2 \psi + I_1I_0 \,, \quad
%	\frac{\omega^2R_0^2}{2kT}=M^2 \,,
%\end{equation}
%where the new condition on $II'$ makes the function $I$ linear in $\psi$. In particular, Eqs.12, 13 are the same while Eq.14 now becomes:
%\begin{align}
%	\laps{\psi} + I_1^2 \psi = -R^2 P_1 e^{M^2R^2} - I_1I_0 \,, \\
%	P\tb{T}(\psi) = 2P_1\psi\,,\quad P(\psi,R)=2P_1\psi e^{M^2R^2} \,, \label{eqpressure} \\
%	I(\psi) = I_1 \psi + I_0 \,,\quad \omega(\psi) = \frac{M}{R_0}\sqrt{2kT(\psi)} \,, \label{eqi}
%\end{align}

%%%%%%%%%%%%%%%%%%%%%%%%%%%%%%%%%%
%%	sol. 1 poly		%%%%%%%%%%%%%%%%%%%%%%
%%%%%%%%%%%%%%%%%%%%%%%%%%%%%%%%%%
\section{polynomial solution}\label{sec-poly}
Concerning the solution of $\laps{\psi\tb{H}}=0$, the usual strategy is to employ separation of variables and assume an expression like $\psi\tb{H}\sim f(R)g(Z)$. The linearity of the equation allows to consider a sum of any number of such terms.
For intance, many authors consider polynomials in the $Z$ variable, and in the special case of up-down symmetry, corresponding to even power only, the following representation can be used \cite{zheng1996}:
\begin{equation}\label{eqhomsum}
\psi\tb{H}=\sum_{n=0,2,...}\sum_{k=0}^{n/2} f_{n,k}(R) Z^{n-2k}\,.
\end{equation}
It can be verified by substitution that the functions $f_{n,k}(R)$ are given recursively by the relations
\begin{align}
&(\partial_R^2-\partial_R/R)f_{n,0} = 0 \,, \nonumber \\
&(\partial_R^2-\partial_R/R)f_{n,k}=-(n-2k+1)(n-2k+2)f_{n,k-1}\,.
\end{align}
This representation of the homogeneous solution in terms of the lowest even powers of $Z$ is suitable for describing up--down symmetric configurations in terms of the minimum number of parameters. For our first analysis, we consider \eref{eqhomsum} truncated at a maximum index $n=4$, and further simplified setting some integration constants to 0 to get rid of terms $\propto\ln(R)$, resulting in the following expression:
\begin{align}\label{eq-hompol}
\psi\tb{H}& = C_0+C_2R^2+C_4(R^4-4R^2Z^2) \nonumber\\
&+C_6(R^6-12R^4Z^2+8R^2Z^4) \nonumber\\
&+C_8\left( R^8-24R^6Z^2+48R^4Z^4-64R^2Z^6/5 \right)   \,,
\end{align}
where the constants $C_i$, $i=0,2,4,6,8$, offer enough freedom to fix the plasma minor radius $a$, the triangularity $\delta$, the elongation $\kappa$ and the boundary curvature at the outermost point $c$, via the conditions:
\begin{align}
\psi(1-a,0) = 0 \,, \label{eqbound1}\\
\psi(1+a,0) = 0 \,, \label{eqbound2}\\
\psi(1-a\delta,a\kappa) = 0 \,, \label{eqbound3}\\
\frac{\partial\psi}{\partial R}(1-a\delta,a\kappa) = 0 \,, \label{eqbound4}\\
\frac{\partial^2\psi}{\partial Z^2}\bigg/\frac{\partial\psi}{\partial R}(1+a,0)=c\,.\label{eqbound}
\end{align}
Concerning the constants $P_1,I_1,I_0$ contained in Eqs.(\ref{eqi}) and (\ref{eqpart}), we set their values according to the desired plasma poloidal beta $\beta\tb{p}$, current $I\tb{p}$ and toroidal magnetic field on axis, given by:
\begin{align}
&\beta\tb{pol} = \frac{\int P\dd{s}}{\int\dd{s}} = \frac{P_1}{S} \int R\frac{\psi(R,Z) e^{M^2R^2}}{B^2}\dd{s} \,,\quad \label{eq-int1}\\
&I\tb{p} = \int J_\phi\dd{s} = \int R P_1 e^{M^2R^2} + \frac{I_1}{R} \dd{s}\,, \label{eq-int2}\\
&B_\phi(R_0,0) = \frac{\sqrt{2I_1\psi(R_0,0)+I_0}}{R_0}\,. \label{eq-int3}
\end{align}
The integrals in the above equations are performed over the confined plasma region inside the magnetic separatrix, defined by $\psi(R,Z)=0$ and corresponding to the boundary between closed and open magnetic lines. We observe here that, even though in the presence of plasma motion the pressure is not constant on magnetic surfaces, like in the static case, according to \eref{eqpressure} we still have $P=0$ on the separatrix just defined. Hence the magnetic and matter boundaries of the plasma coincide.
The same region can also be described by the points $(R,Z)\in\{1-a,1+a\}\times\{-Z\tb{m}(R),Z\tb{m}(R)\}$, with $\psi(R,Z\tb{m}(R))=0$. The function $Z\tb{m}(R)$ can be calculated explicitly, describing the plasma upper boundary (or lower, with minus sign) in terms of $a$, $\delta$, $\kappa$, $c$, $P_1$, $I_1$ and $M$, however we omit its cumbersome expression for brevity. 
\begin{figure}
\centering
\includegraphics[width=0.55\columnwidth]{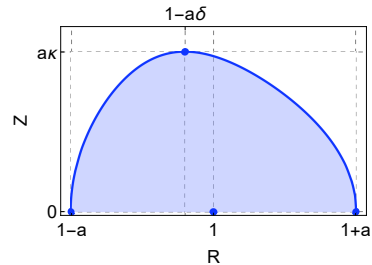}
\caption{Analytic plasma shape defined as the curve $\psi(R,Z)=0$, parametrized through the constants $a$, $\delta$, $\kappa$ and $c$, for some arbitrary values of the physical constants $P_1$, $I_1$ and $M$. \label{fig-shape}}
\end{figure}
In general, closed--form expressions for the integrals cannot be found, hence we resort to standard numerical recipes for their calculation. In the practical implementation, we find that a simple \emph{guess and check} strategy leads to satisfying results after a single iteration.

\subsection*{Study of rotation influence}
\begin{figure}
\centering
%[for twocolumn]
\includegraphics[width=0.44\columnwidth]{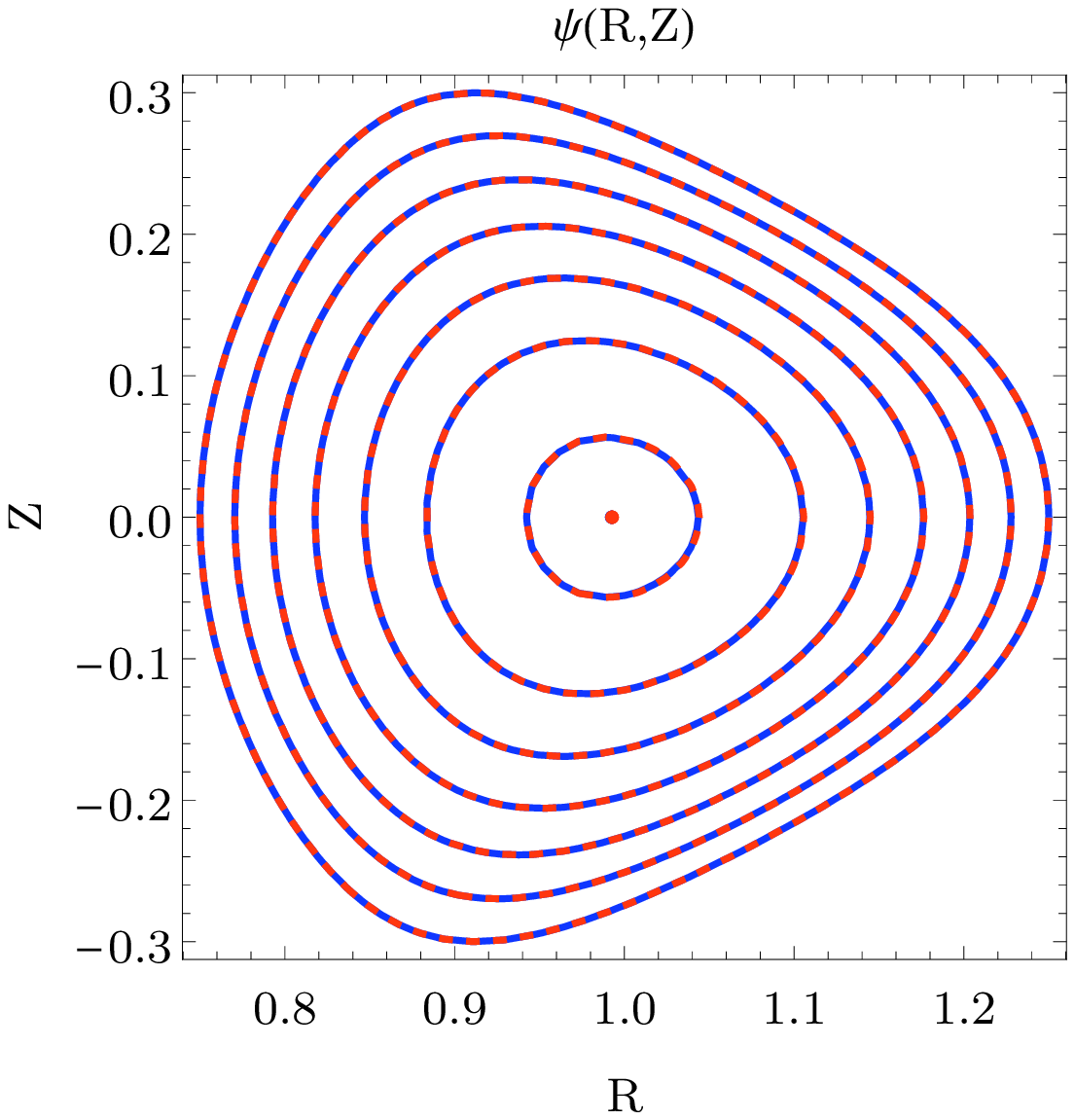}
\includegraphics[width=0.44\columnwidth]{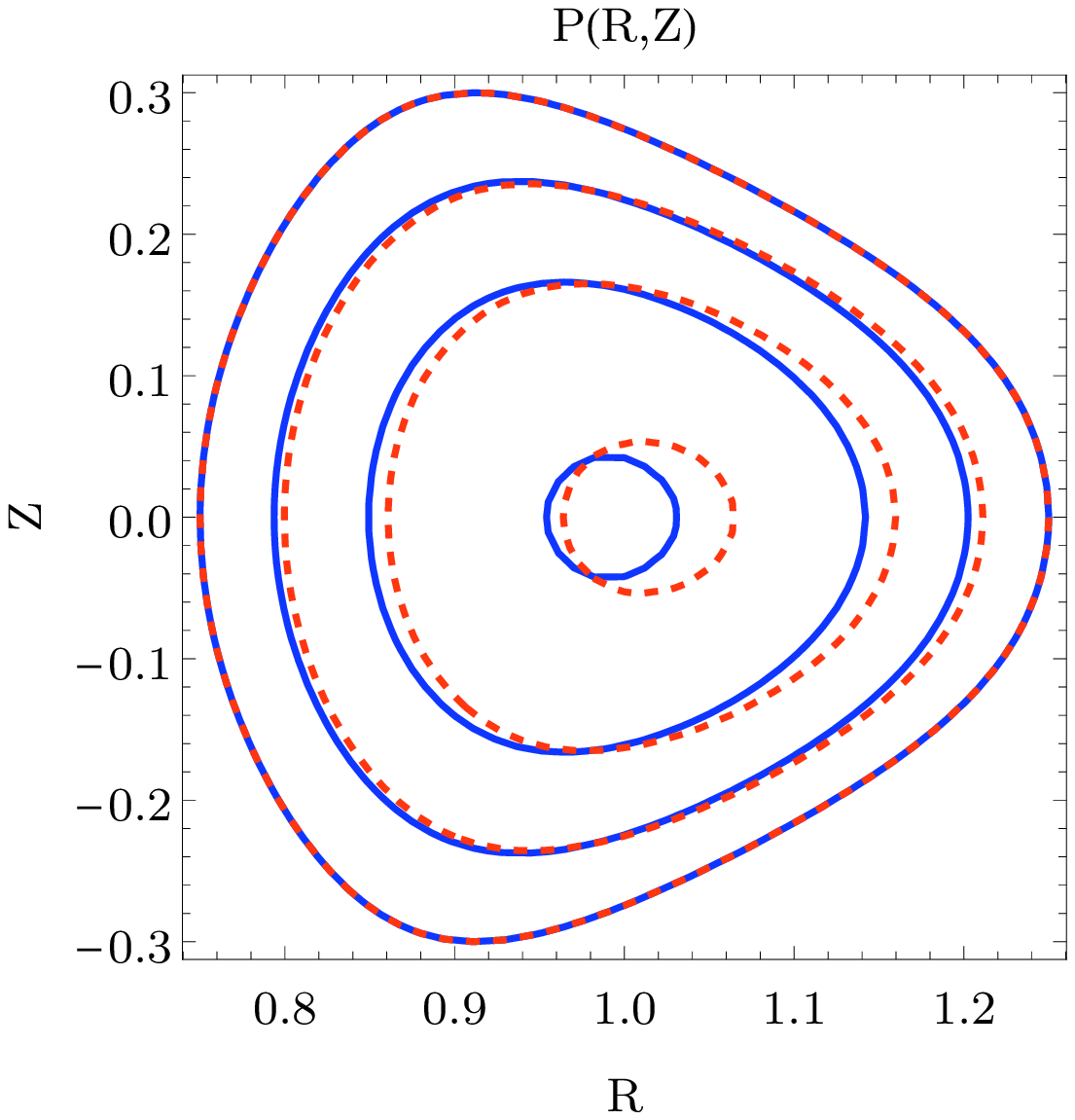}\\
\vspace{0.3cm}
\includegraphics[width=0.44\columnwidth]{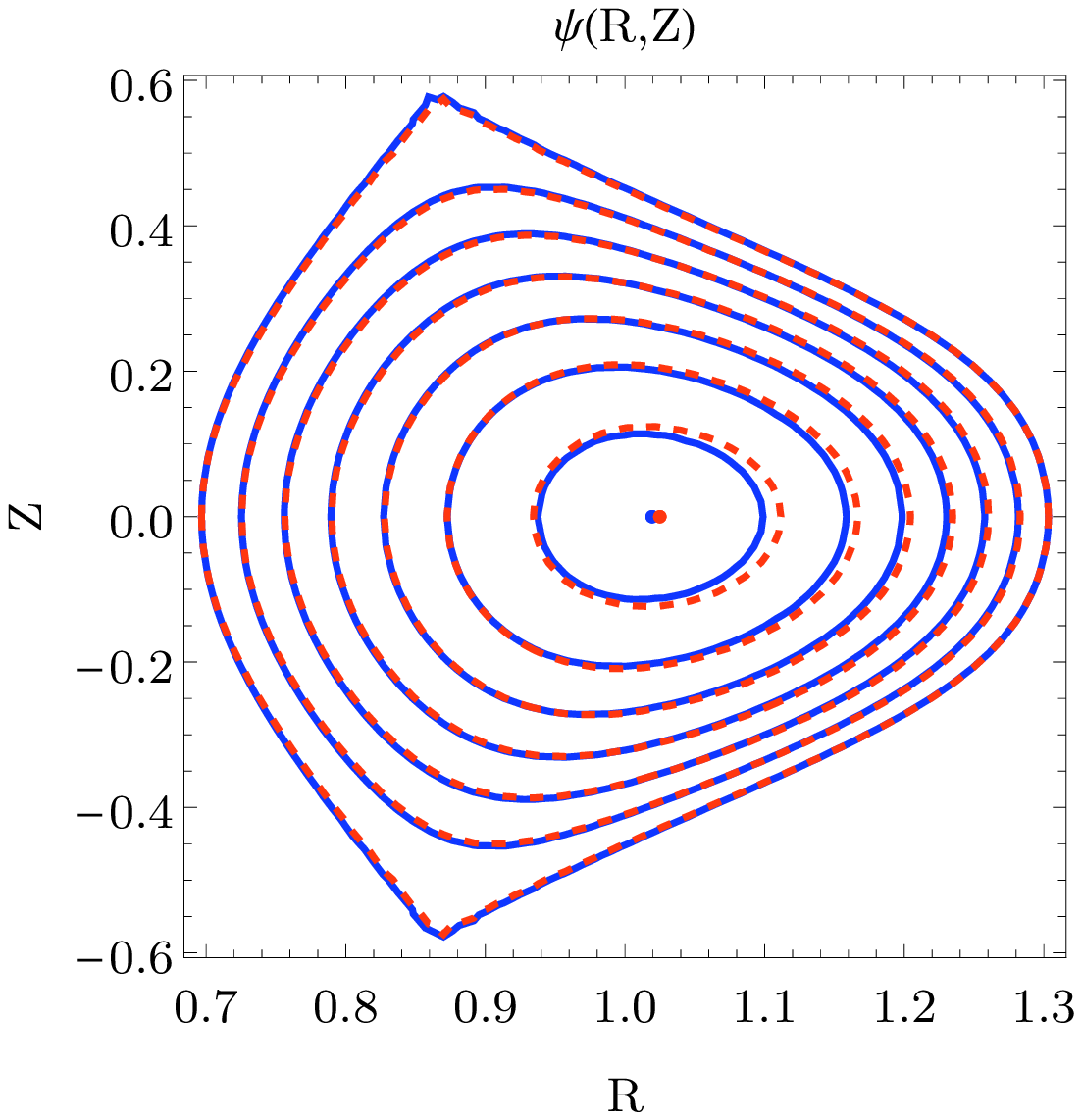}
\includegraphics[width=0.44\columnwidth]{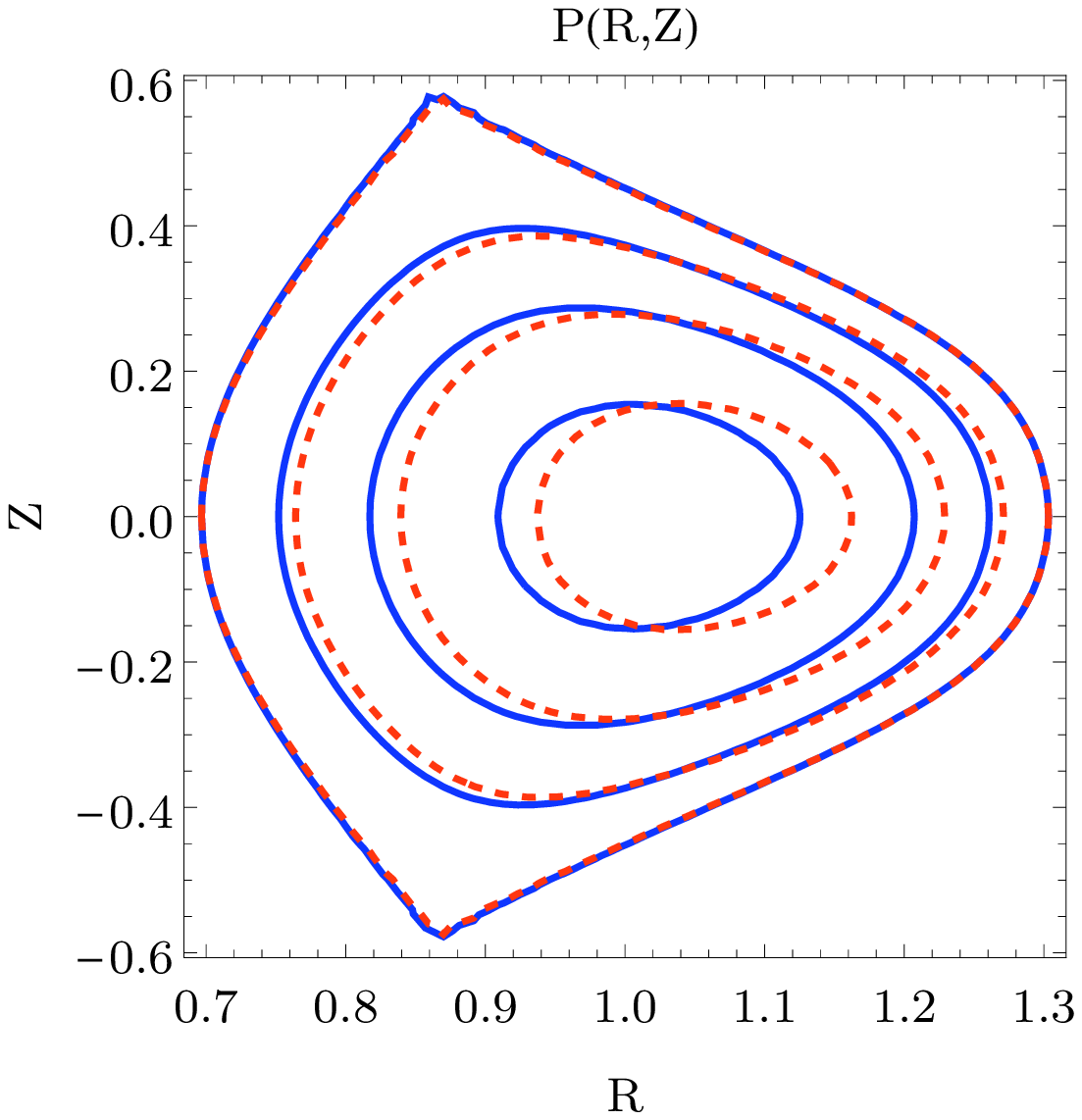}\\
\vspace{0.3cm}
\includegraphics[width=0.44\columnwidth]{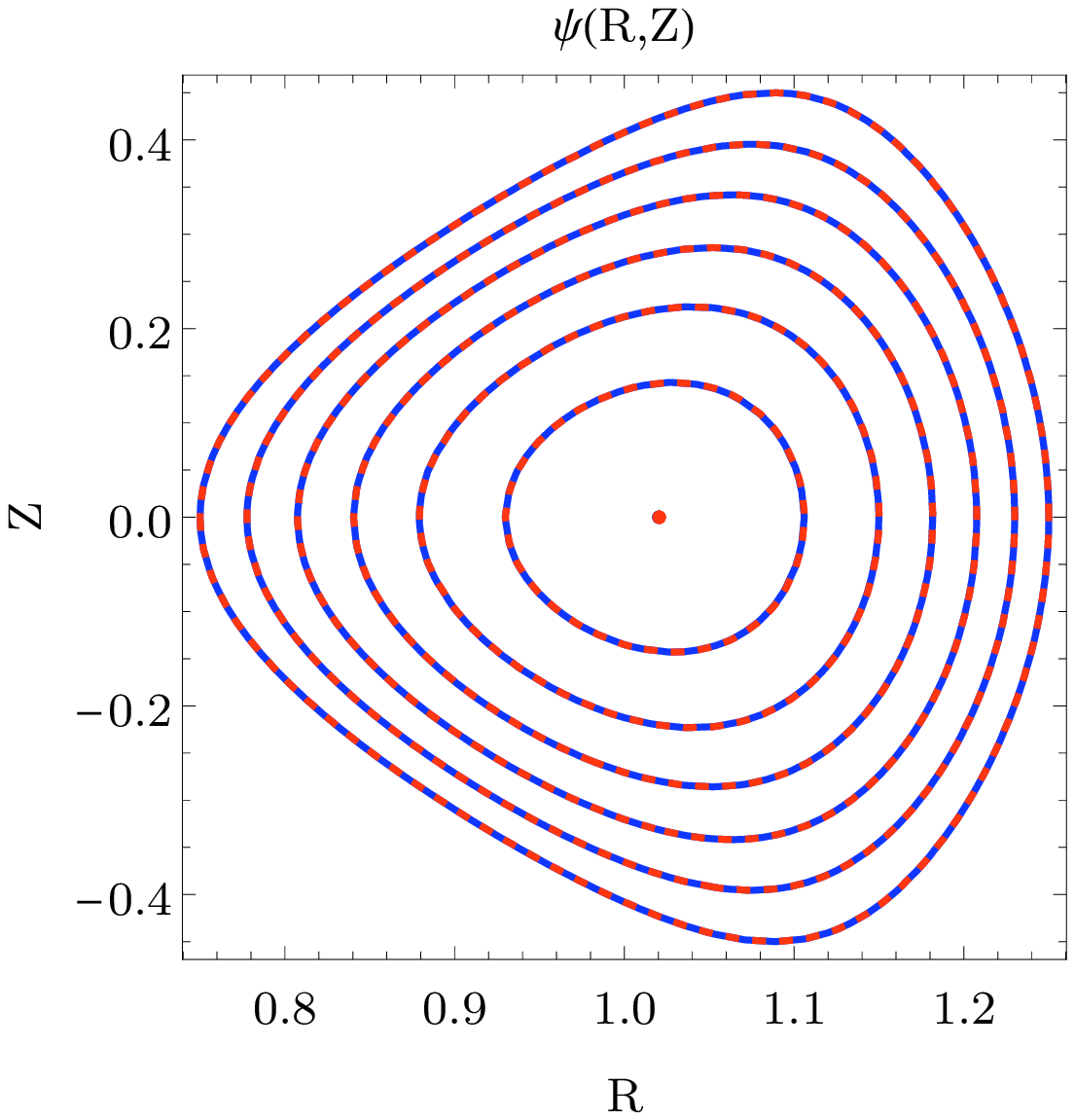}
\includegraphics[width=0.44\columnwidth]{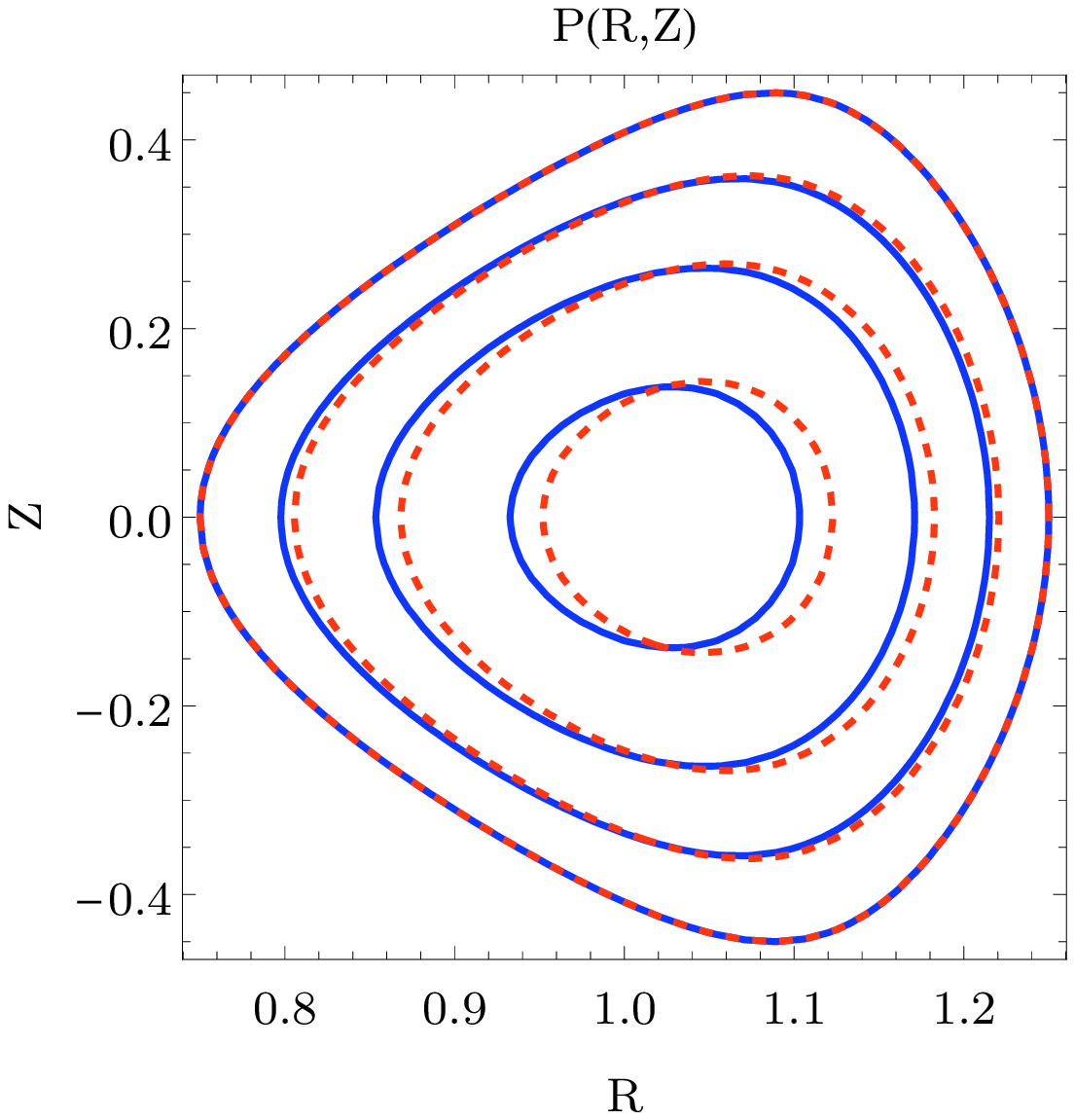}
\caption{Contours of constant magnetic flux $\psi$ (left) and pressure $P$ (right) in the (R,Z) plane, calculated for the $M=0$ (solid) and $M=0.3$ (dashed) cases and corresponding to the parameters reported in Table 1.\label{fig-polyprofs}}
\end{figure}

We note that the rotation velocity is treated as a free parameter so far, via the constant $M$, with $M=0$ corresponding to the static plasma case. Within this framework, we are able to evaluate its direct impact on the other equilibrium features.
In Fig.\ref{fig-polyprofs} we show the capabilities of solution Eq.(\ref{eq-hompol}) of reproducing plasma shapes with different values of the parameters, as in Table \ref{tab-1}.

% TABLE %%%%%%%%%%%%%%	
{\begin{table}[h]
\centering
\caption{Values of the minor radius $a$, triangularity $\delta$, elongation $\kappa$ and curvature $c$ for the plasma configurations of Fig.\ref{fig-shape}.\label{tab-1}}
\begin{tabularx}{\linewidth}{@{}lYYYY@{}}
\toprule
 & $a$ & $\delta$ & $\kappa$ & $c$ \\
\midrule
first row & 0.25 & 0.35 & 1.20 & 5 \\
second row & 0.30 & 0.45 & 1.92 & 2 \\
third row & 0.25 & -0.35 & 1.80 & 1 \\
\bottomrule
\end{tabularx}
\end{table}}
% ENDTABLE %%%%%%%%%%%%%
With regard to the second row, the pointy shape of the profile at its top and bottom suggests the presence of x-points. However, it must be noted that the solution used here has not enough free constants to impose the proper null condition on the magnetic field at a desired location; the x-points can only emerge at a certain location, fixed by parameters $\delta$ and $\kappa$, for specific choices of $a$ and $c$. We will see in the next Section how to address this shortcoming.

\begin{figure}
\centering
\includegraphics[width=0.74\columnwidth]{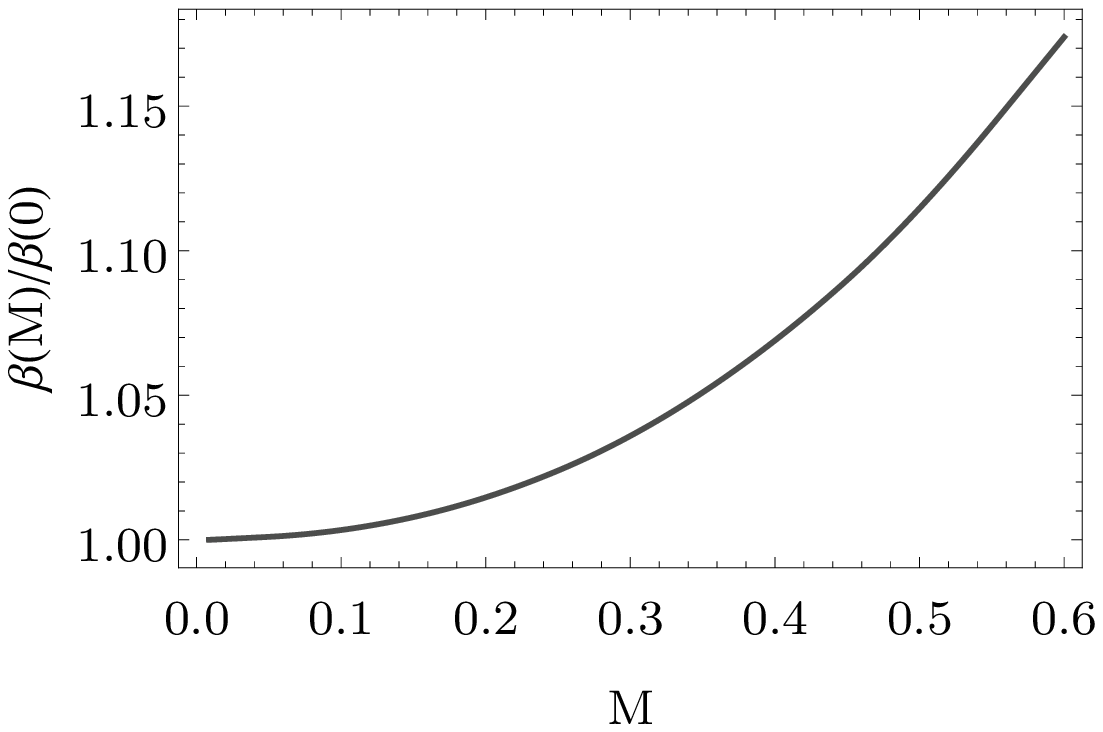} \\
\vspace{0.3cm}
\includegraphics[width=0.74\columnwidth]{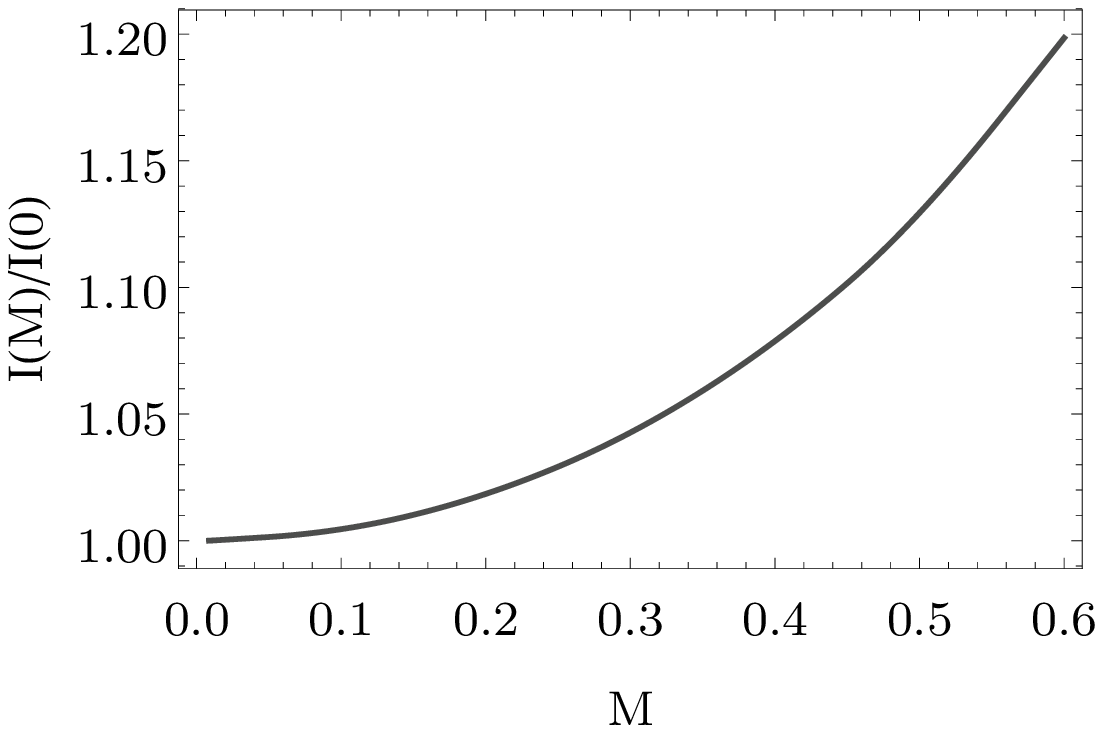} \\
\vspace{0.3cm}
\includegraphics[width=0.74\columnwidth]{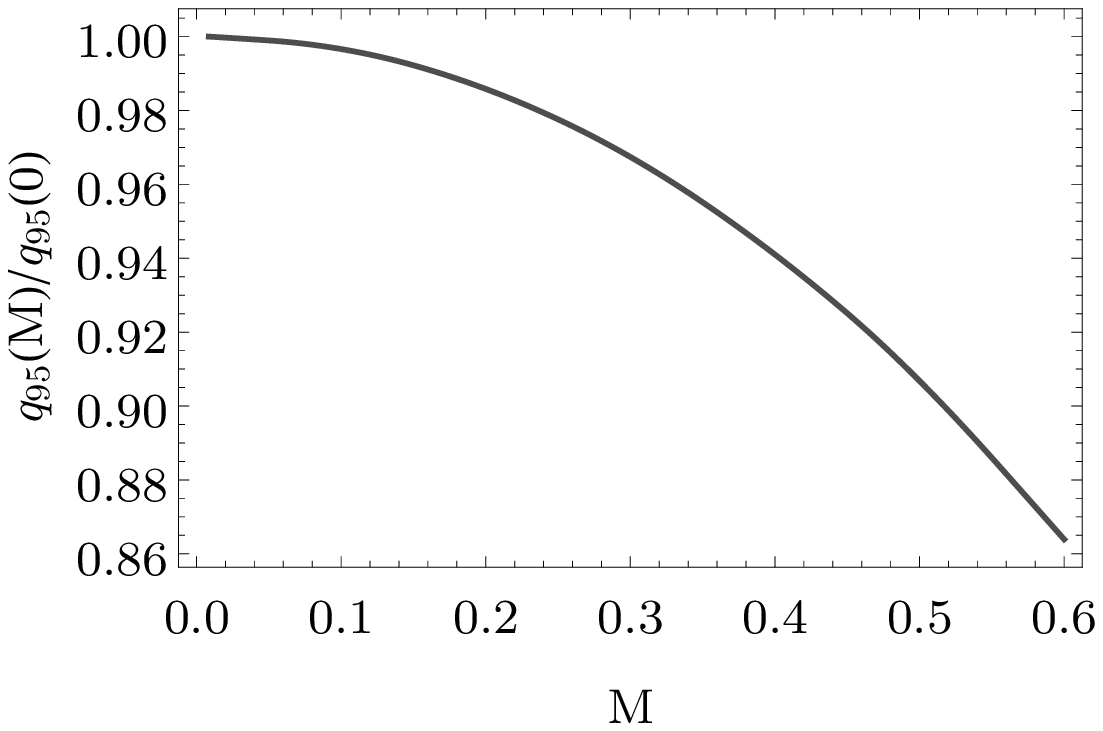} \\
\caption{Variation of poloidal beta, plasma current and safety factor at 95\% plasma volume, with respect to the parameter $M$, each normalized by their respective value in the static case. The other parameters of the configuration are kept fixed and correspond to Table 1, second row.\label{fig-profs}}
\end{figure}

Fig.\ref{fig-profs} shows the dependence on $M$ of $\beta\tb{pol}$, $I\tb{p}$ and $q_{95}$, the safety factor at 95\% plasma volume, normalized to their respective values in the static case. The other parameters used for the fit correspond to the second row scenario of Table 1. While the entity of the variations differ for other choices of the parameters, however, the general qualitative behaviour is consistently that of an enhancement of both $\beta\tb{pol}$ and $I\tb{p}$, while the safety factor is suppressed. In all the considered cases this has never resulted in breaking the Kruskal-Shafranov stability condition $q>1$ over the whole plasma profile, even for unrealistically high rotation velocities. 

These results suggest that, when modeling real plasma equilibria using a static analytical solution or numerical code, the errors commited can get increasingly large with plasma rotation. The evaluation of $M$ can thus give quantitative insight on the necessity to employ an exact solution or equilibrium solver which take the plasma rotation into account.
%%%%%%%%% END TWO FIGS %%%

%A different strategy consists in keeping some plasma features fixed at different values of $M$, by recalculating the equilibrium each time, and looking for variations in other observables, such as the plasma shape or the position of the magnetic axis. In this way, we focus on the impact of including plasma rotation in the equilibrium reconstruction when other parameters are known. This could be relevant in realistic situations where direct measurements are available.

%Figure W shows the coefficients $A_1$ and $A_2$ needed at different values of $M$ in order to keep the plasma current and beta parameter fixed. It emerges how $A_2$ is almost constant, while $A_1$ decreases. Recalling their relation to the toroidal component of the magnetic field and the pressure, respectively,CIT EQUAZIONI we conclude that in this case the introduction of rotation leads to a suppression of the pressure profile inside the plasma. However, this should not be interpreted as a suppression of the beta parameter itself, since here it is being held constant. 

%%%%%%%%%%%%%%%%%%%%%%%%%%%%%%%%%%
%%	sol. 2 bessel		%%%%%%%%%%%%%%%%%%%%%%
%%%%%%%%%%%%%%%%%%%%%%%%%%%%%%%%%%
\section{general solution}\label{sec-bessel}
%The solution \eref{eqhompol} studied in the previous section is easy to implement, requiring only few contraints in order to obtain the equilibrium plasma profile. However, such solution is not suitable when the plasma boundary curve is known, either analytically or numerically, and an accurate fit is needed. Namely, the position of x-points, \ie, boundary points with vanishing magnetic field gradient, can only be fixed correctly at the cost of neglecting other constraints. 
Suppose that the plasma boundary curve is known, either analytically or numerically, and one wants an accurate fit reproducing its shape. The polynomial solution Eq.(\ref{eq-hompol}) studied in the previous section is easy to implement, requiring only few contraints in order to obtain the equilibrium, but it evidently fails in such situations. As discussed above, the position of x-points, \ie boundary points with vanishing magnetic field gradient, can be fixed exactly only at the cost of neglecting other constraints. This issue could be overcome by extending the expression in Eq.(\ref{eq-hompol}) up to a suitable higher power of $Z$, thus generating new free constants in the solution. However, we show here that the general solution of the homogeneous equation $\laps{\psi\tb{H}}=0$ can be written in a compact form, without any truncation, and that it can be used to solve this kind of problem.

To find the general solution, we express $\psi\tb{H}$ as a Fourier transform in the $Z$ variable:
\begin{equation}\label{eqpsifou}
\psi\tb{H}(R,Z)=\int_{-\infty}^{\infty}\chi(R,k)\me^{ikZ}\dd{k}.
\end{equation}
Its reality is ensured by the condition $\chi(R,-k)=\overline{\chi(R,k)}$. Plugging \eref{eqpsifou} into $\laps{\psi\tb{H}}=0$, we obtain an ordinary differential equation in the variable $R$ for each $k$. By making the change of variables $x_k=|k|R$ and $\chi(R,k)=R\epsilon(x_k,k)$ (for $k\neq0$), it is easy to verify that:
\begin{equation}
x_k^2\epsilon(x_k,k)''+x_k\epsilon(x_k,k)'-\left(1+x_k^2\right)\epsilon(x_k,k)=0\;,
\end{equation}
where the prime denotes differentiation with respect to $x_k$. This is known as Bessel's modified equation with index 1, and its solution is readily available in mathematical literature:
\begin{equation}
\epsilon(x_k,k)=a_{k}I_1(x_k)+b_{k}K_1(x_k)  \,,
%\epsilon(x_k,k)=a_{k}I_1(x_k)+b_{k}K_1(x_k)  \;\;\text{or}\;\;  m_kJ_1(-i x_k)+n_kY_1(-i x_k)\,,
\end{equation}
where $a_{k}, b_{k}$ %, m_k,n_k$ 
are functions of $k$. By substitution back into \eref{eqpsifou}, we obtain:
\begin{equation}\label{eqhomfou}
\psi\tb{H}(R,Z)=R\int_{-\infty}^{\infty} \left[ a_k I_1(|k|R) + b_k K_1(|k|R) \right] \me^{ikZ}\dd{k}\,,
\end{equation}
which is the general solution of the homogenous problem.

%This solution allows us to precisely reconstruct the plasma profile once the separatrix shape is known, either as an analitical or numerical curve. To do so, we represent the functions $a_k$ and $b_k$ as a sum of delta functions, selecting a set of wavenumers $k_i$ with constant weights $a_{k_i}$, $b_{k_i}$. we extract a suitable number of boundary points from the profile where we impose $\psi=0$, as in \emphr{FigureX}. Then, we represent the functions $a_k$ and $b_k$ as a sum of delta functions, selecting a set of wavenumbers $k_i$ which are chosen in order to match 
%%%%%%%%%%%%%%%%%%%%%%%%%%%%%%%%%%%%
This expression can be adapted to a given scenario by imposing specific boundary conditions. In this respect, for the sake of simplicity, we represent the functions $a_k,\,b_k$ as a sum of sufficiently narrow gaussians (\ie delta functions), centered around arbitrarily given wave vectors $k_i$ and weighted by amplitudes $\bar{a}_i,\,\bar{b}_i$, thus obtaining:
\begin{equation}
\psi\tb{H}(R,Z) = R \sum_{i=1}^N \left[\bar{a}_i I_1(R|k_i|) + \bar{b}_i K_i(R|k_i|) \right] \cos(k_i Z)\,,
\end{equation}
where the term $\cos(k_iZ)$ is the reduction of the complex exponential to the real, up-down symmetric case.
Then, a given set of points $\{r_l,z_l\}$ lying along the boundary curve of the addressed plasma configuration generates an associated set of algebraic equations of the form $\psi(r_l,z_l)=0$, which can be solved to determine the arbitrary constants.
%%%%%%%%%%%%%%%%%%%%%%%%%%%%%%%%%%%%
\subsection*{DTT double-null configuration}
We illustrate this procedure in the practical case of the double-null plasma scenario predicted for the upcoming DTT experiment. Its main parameters are reported in Table \ref{tab-2}, and are available in Ref.\cite{dttgreen} along with the predicted separatrix shape. 
% TABLE %%%%%%%%%%%%%%	
{\begin{table}[h]
\centering
\caption{Main plasma parameters of the DTT double-null scenario, taken from \cite{dttgreen}.\label{tab-2}}
\begin{tabularx}{\linewidth}{@{}YYYYYYY@{}}
\toprule
$I\tb{p}\,$(MA) & $\beta\tb{pol}$ & $q_{95}$ & $R_0\,$(m) & $a\,$(m) & $\delta$ & $\kappa$ \\
\midrule
5.00 & 0.43 & 2.80 & 2.11 & 0.64 & 0.45 & 1.92 \\
\bottomrule
\end{tabularx}
\end{table}}
We proceed as follows: firstly, we model the desired separatrix as an analytic curve, using a piecewise rational expression (e.g. quadratic). Secondly, we extract a set of boundary points chosen at random but equally distributed around the plasma region. Thirdly, the set of wavenumbers $k_i$ is chosen as an equally distributed grid of values close to the scale length of the configuration, estimated as $\pi/(a\kappa)$.
The solution of the resulting set of algebraic equations gives the constants $\bar{a}_i,\,\bar{b}_i$ as functions of $P_1$, $I_1$ and $M$. The former two are still obtained according to Eqs.(\ref{eq-int1}), (\ref{eq-int2}) and (\ref{eq-int3}), and we can study the behaviour of the equilibrium for different values of $M$.

\begin{figure}
\centering
%[for twocolumn]
\includegraphics[width=0.44\columnwidth]{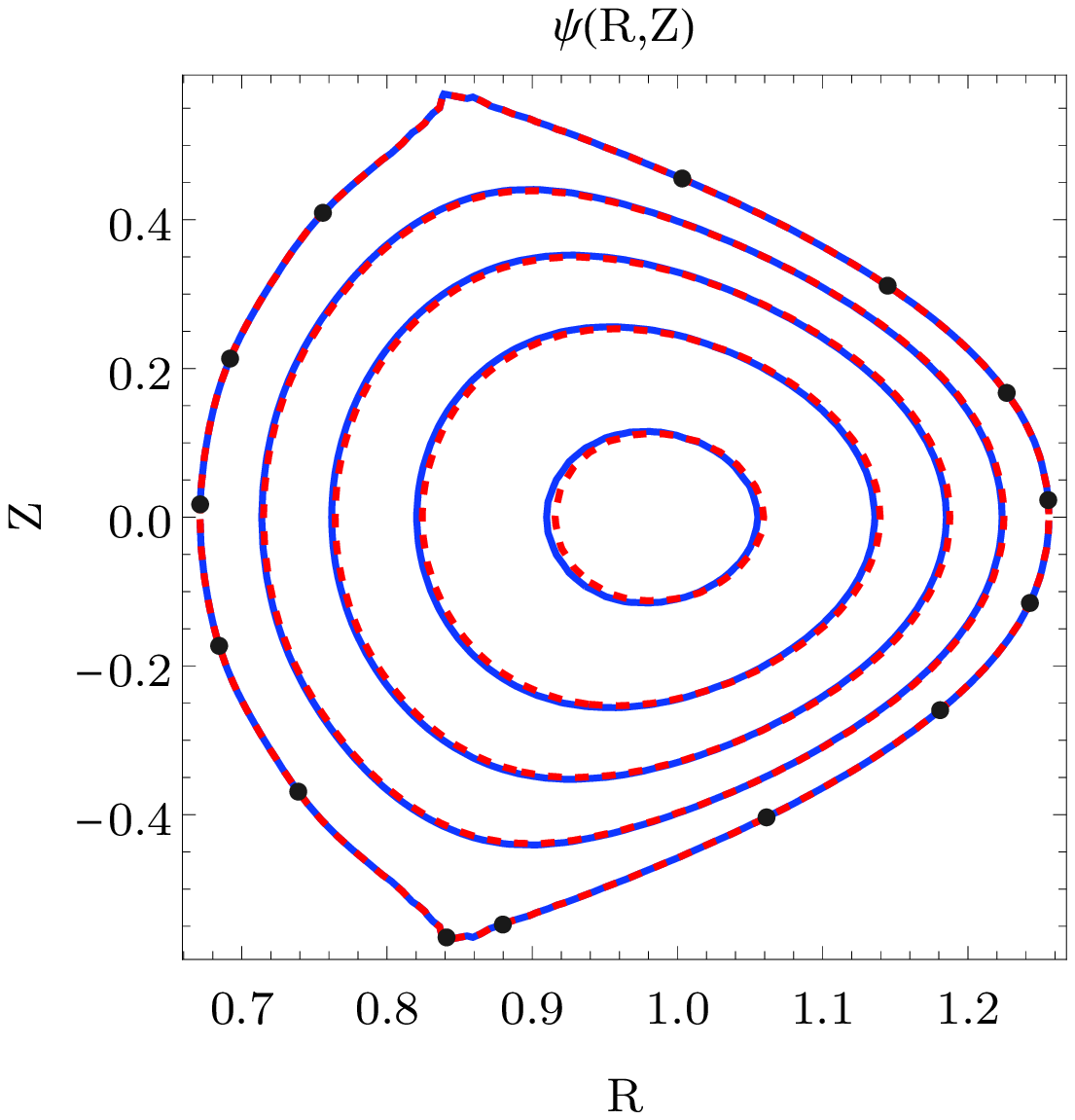}\hfill
\includegraphics[width=0.44\columnwidth]{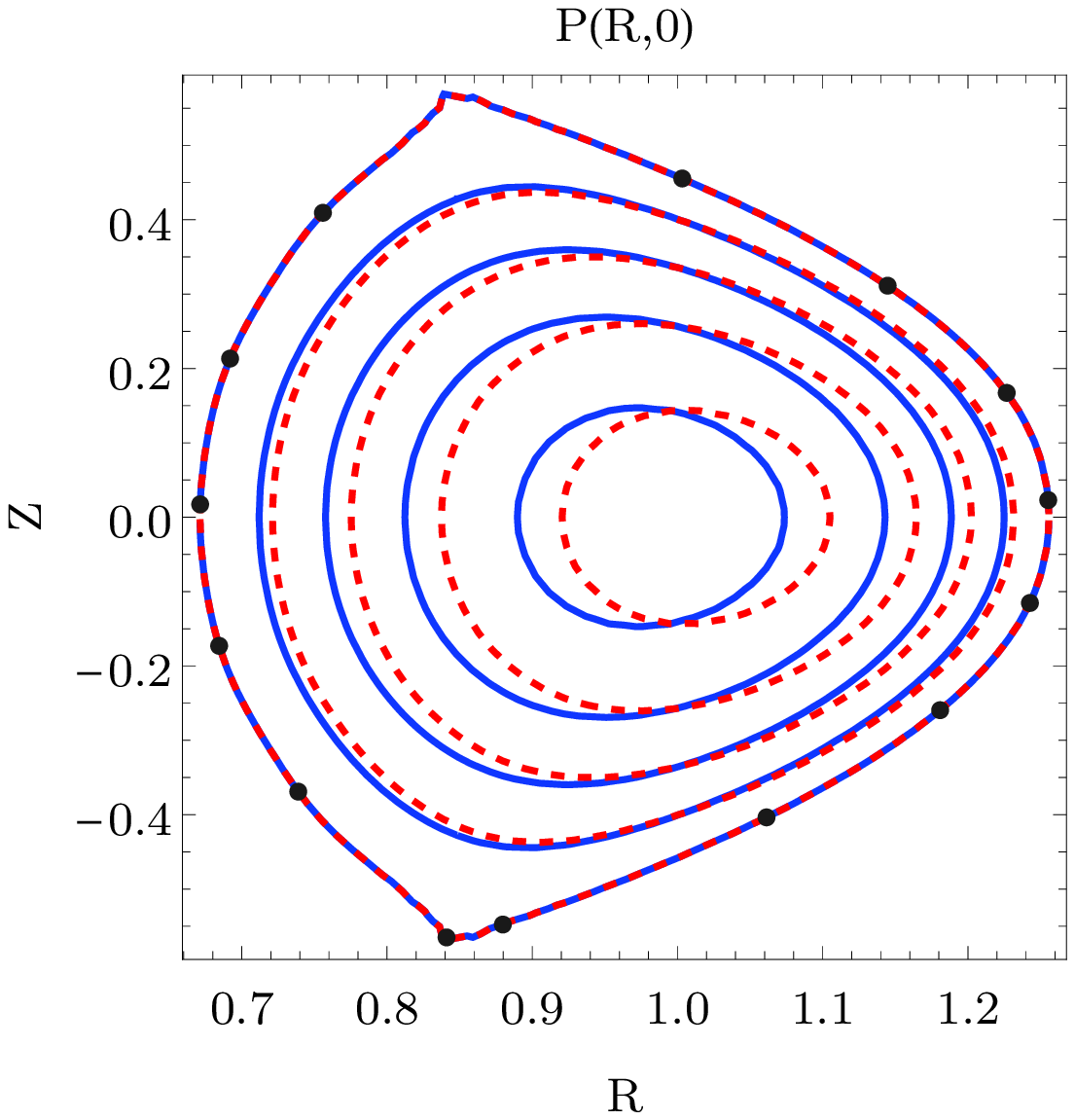}
\caption{Contours of constant $\psi$ (left) and $P$ (right) in the (R,Z) plane for the DTT double-null plasma, in the static case ($M=0$, solid blue) and rotating case ($M=0.6$, dashed red).\label{fig-p}}
\end{figure}
Fig.\ref{fig-p} shows the fitted magnetic configuration and the curves of constant pressure, where we highlighted the correspondence between the set of boundary fitting points and the obtained separatrix. In the presence of rotation, the qualitative behaviour of the plasma is still that of an outward shift of magnetic and pressure lines, while the separatrix is kept fixed by the imposed constraints and has no major modifications. 
\begin{figure}
\centering
\includegraphics[width=0.64\linewidth]{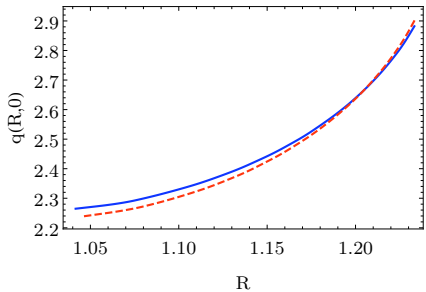}
\caption{Safety factor profile of the DTT double-null scenario in the static case ($M=0$, solid blue) and rotating case ($M=0.6$, dashed red).\label{fig-q}}
\end{figure}
Concerning the safety factor profile, plotted in Fig.\ref{fig-q}, we predict a slight suppression in the core region, while closer to the plasma boundary $q$ actually increases, contrary to the general behaviour observed in the previous section using solution (\ref{eq-hompol}).
\begin{figure}
\centering
\includegraphics[width=0.44\linewidth]{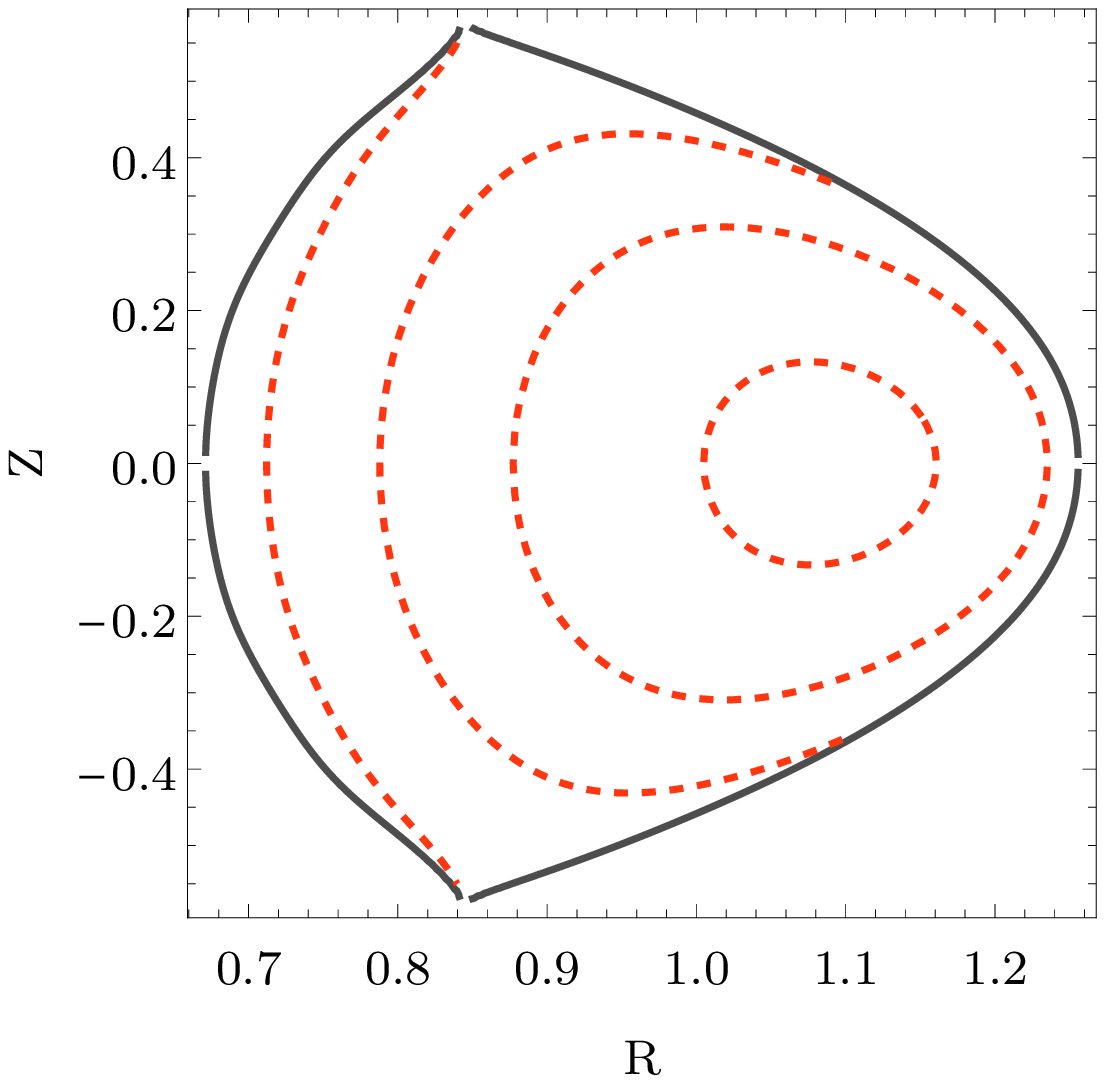}\hfill
\includegraphics[width=0.44\linewidth]{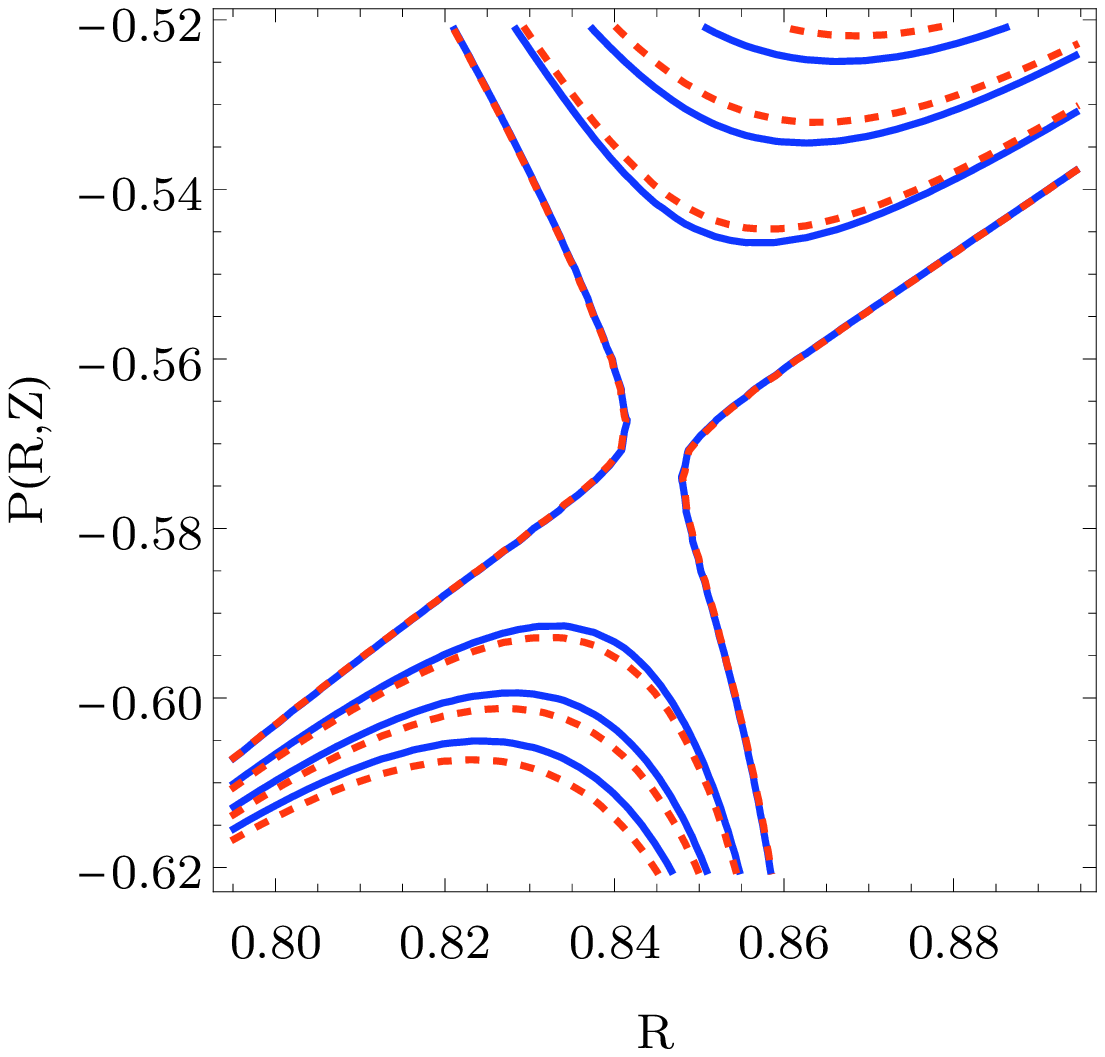}
\caption{Contours of constant toroidal velocity $\omega R$ over the whole configuration (left), and contours of constant pressure in the vicinity of the x-point (right, $M=0$ solid blue, $M=0.6$ dashed red).\label{fig-rot-x}}
\end{figure}
Finally, we can plot the curves of constant toroidal speed $\omega(\psi) R$ by assuming a simple form for the temperature, taken as $T(\psi) = T\tb{edge} + \frac{\psi}{\psi\tb{axis}}(T\tb{core}-T\tb{edge})$, with the temperature values according to \cite{dttgreen}. The result is shown in Fig.\ref{fig-rot-x}, along with the morphology of pressure lines in the vicinity of the x-point. In this formalism, we don't expect serious modifications to the shape of the plasma in this region, having imposed our constraints along the separatrix itself. However, observing how the pressure (and its gradient) are suppressed in the presence of rotation might provide useful information when considering transport dynamics. For example, it is common practice to feed equilibrium data obtained from a given solver into a separate code which simulates particle transport. A scenario in which the parameter $M$ is measured to be consistently far from 0, giving rise to noticeable modifications of the equilibrium, would need to take plasma rotation into account. Of course, the present analysis is aimed at providing a simple semi-analytical tool to gain insight in this direction, while accurate equilibrium solvers with plasma flow should be used for more elaborate analysis (e.g. \cite{gbmk2004}).

%%%%%%%%%%%%%%%%%%%%%%%%%%%%%%%%%%
%%	conclusion		%%%%%%%%%%%%%%%%%%%%%%
%%%%%%%%%%%%%%%%%%%%%%%%%%%%%%%%%%
\section{concluding remarks}
In this work, we studied the equilibrium of an axisymmetric plasma in the presence of rotation along the toroidal direction. After recalling the mathematical basic formalism, we adopted suitable assumptions on the arbitrary functions in order to obtain analytic plasma profiles, with enough freedom to represent a variety of plasma settings. In particular, the polynomial expression of section \ref{sec-poly} requires to fix only few basic plasma parameters, such as the minor radius and the triangularity. This simplicity allows to find analytical expressions for the plasma separatrix and to deal with a variety of scenarios, e.g. double-null and negative triangularity. 
Then, in section \ref{sec-bessel}, we presented the general solution of the considered problem, and illustrated a suitable fitting procedure when dealing with a known plasma separatrix (either analytically or numerically).
As a practical implementation of this framework, we studied the double-null plasma scenario proposed for the upcoming Italian experiment DTT, estimating the impact of plasma rotation on the equilibrium properties and highlighting some points of interest such as the modification of the plasma pressure gradient morphology in the vicinity of the x-point, with possible effects on particle transport dynamics in that region.

Of course, the analysis performed here has the merit of simplicity due to its analytic nature, but needs to be confirmed by more detailed numerical studies when dealing with more realistic situations. Moreover, many physical constraints here neglected (e.g., the specifics of the given tokamak magnetic coils, or its current drive mechanism) would need to be taken into account. Nevertheless, the two approaches of section \ref{sec-poly} (reduced) and \ref{sec-bessel} (general) agree on the qualitative behaviour of the plasma parameters as functions of the rotation velocity, hence they can both be used as quick investigative tools concerning the introduction of toroidal rotation in tokamak plasma equilibria. In experimental situations, the parameter $M$ can be estimated from Eq.(14) providing direct measurements of ion rotation speed and temperature, e.g. through diagnostics like charge exchange recombination spectroscopy \cite{fonck}. Depending on the value of $M$, quantitative estimates on the relevance of plasma rotation can be argued by the methods outlined here.

%%%%%%%%%%%%%%%%%%%%%%%%%%%%%%%%%%
%%	bibliography		%%%%%%%%%%%%%%%%%%%%%%
%%%%%%%%%%%%%%%%%%%%%%%%%%%%%%%%%%

%\bibliography{mybib}

\begin{thebibliography}{99}
\bibitem{wesson}
J. Wesson, \emph{Tokamaks} (Oxford University Press) 1997.

\bibitem{biskamp}
D. Biskamp, \emph{Nonlinear magnetohydrodynamics} (Cambridge University Press) 1993.

\bibitem{shafranov}
V.D. Shafranov, \emph{Rev. Plasma Phys.} \textbf{2}, 103 (1966).

\bibitem{hassam93}
A.B. Hassam, T.M. Antonsen Jr., J.F. Drake, P.N. Guzdar, C.S. Liu, D.R. McCarthy, and F.L. Waelbroeck, \emph{Phys. Fluids B} \textbf{5}, 2519--2524 (1993) 
%"Spontaneous and driven perpendicular rotation in tokamaks", https://doi.org/10.1063/1.860738

\bibitem{duval07}
B.P. Duval, A. Bortolon, A. Karpushov, R.A. Pitts, A. Pochelon and A. Scarabosio, \emph{PPCF} \textbf{49}, B195--B209 (2007)

\bibitem{karpushov17}
A. Karpushov et al., \emph{Fusion Eng. Des.} \textbf{123}, 468--472 (2017).

\bibitem{mp1980}
E.K. Maschke and H. Perrin, Plasma Physics {\bf 22}, 6, 579--594 (1980)

\bibitem{hameiri83}
E. Hameiri, \emph{Phys. Fluids} \textbf{26}, 230--237 (1983)

\bibitem{ogilvie97}
G.I. Ogilvie, \emph{Mon. Not. R. Astron. Soc.} \textbf{288}, 63--77 (1997)

%\bibitem{landau8}
%L.D. Landau, E.M. Lifshitz, \emph{Electrodynamics of Continuous Media} (Pergamon Press) 1960. 

\bibitem{solo68}
L.S. Solov'ev, \emph{Sov. Phys. JETP} \textbf{26}(2), 400 (1968).

\bibitem{dttgreen}
R. Albanese, F. Crisanti, P. Martin, R. Martone, A. Pizzuto and DTT Contributors, Divertor Tokamak Test Facility, interim design report (2019)

\bibitem{gbmk2004}
L. Guazzotto, R. Betti, J. Manickam and S. Kaye, Physics of Plasmas {\bf 11}, 2, 602--614 (2004)

%\bibitem{mccar99}
%P.J. Mc Carthy, \emph{Physics of Plasmas} \textbf{6}, 3554 (1991).

\bibitem{ferraro}
V.C.A. Ferraro, \emph{Mon. Not. RAS} \textbf{97}, 458 (1937)

%\bibitem{Montani18}
%G. Montani, M. Rizzo, N. Carlevaro, \emph{Phys. Rev. E}, \textbf{97}, 023205 (2018)

\bibitem{guazz21}
L. Guazzotto and J. Freidberg, \emph{JPP} \textbf{87(3)}, 905870305 (2021)

\bibitem{zheng1996}
S. B. Zheng, A. J. Wootton and Emilia R. Solano, Physics of Plasmas {\bf 3}, 3, 1176--1178 (1996)

\bibitem{fonck}
R.J. Fonck, R.J. Goldston, R. Kaita, and D.E. Post, \emph{Applied Physics Letters} \textbf{42}, 239--241 (1983)


\end{thebibliography}

\end{document}